\documentclass[12pt]{amsart}

\usepackage[left=15mm, right=15mm, top=15mm, bottom=15mm]{geometry}

\usepackage{float}
\usepackage[font=small,labelsep=period, figurename={Рис.}]{caption}

\usepackage[english]{babel}
\usepackage[T1]{fontenc}
\usepackage[utf8]{inputenc}

\theoremstyle{plain}
    \newtheorem{theorem}{Theorem}[section]
    \newtheorem{lemma}[theorem]{Lemma}
    \newtheorem{proposition}[theorem]{Proposition}
    \newtheorem{corollary}[theorem]{Corollary}

\theoremstyle{definition}

\usepackage[
    draft = false,
    unicode = true,
    colorlinks = true,
    allcolors = blue,
    hyperfootnotes = true,
    citecolor = red
]{hyperref}
\usepackage{amsfonts,amsmath,amssymb,amscd,amsthm,url}
\usepackage[inline]{enumitem}
\usepackage{graphicx,epsf,afterpage, wrapfig}
\usepackage{soul}
\usepackage{multicol}
\usepackage{array}
\usepackage{braket}
\usepackage{epigraph}
\usepackage{rotating, floatflt}
\usepackage{xstring} 

\usepackage[
backend=biber,
style=numeric-comp,
sorting=none,
giveninits=true
]{biblatex}
\usepackage{csquotes}

\usepackage{xcolor}

\usepackage{ulem}

\usepackage{tikz}
\usetikzlibrary{positioning, calc, intersections, through, arrows, matrix, chains, math}
\usetikzlibrary{arrows.meta}
\usetikzlibrary{shadows}
\usetikzlibrary{graphs}
\usetikzlibrary{graphs.standard}
\usetikzlibrary{patterns}
\usetikzlibrary{decorations.pathreplacing,calligraphy,backgrounds}

\DeclareFieldFormat{usera}{\href{https://arxiv.org/abs/#1}{\texttt{arXiv:#1}}}

\renewbibmacro*{finentry}{\printfield{usera}\newunit\finentry}

\renewbibmacro{in:}{%
  \iffieldundef{usera}
    {\printtext{\bibstring{in}\intitlepunct}}
    {}%
}

\newcommand\norm[1]{\ensuremath{\left\lVert#1\right\rVert}}
\newcommand\abs[1]{\ensuremath{\left\lvert#1\right\rvert}}

\DeclareMathOperator{\Tr}{Tr}

\newcommand{\Acal}{\mathcal{A}}
\newcommand{\Bcal}{\mathcal{B}}
\newcommand{\Ccal}{\mathcal{C}}

\newcommand{\Fcal}{\mathcal{F}}

\newcommand{\Hcal}{\mathcal{H}}
\newcommand{\Ical}{\mathcal{I}}

\newcommand{\Kcal}{\mathcal{K}}

\newcommand{\Mcal}{\mathcal{M}}

\newcommand{\Scal}{\mathcal{S}}
\newcommand{\Tcal}{\mathcal{T}}

\newcommand{\Abf}{{\ensuremath{\mathbf{A}}}}

\newcommand{\Ibf}{{\ensuremath{\mathbf{I}}}}

\newcommand{\Kbf}{{\ensuremath{\mathbf{K}}}}

\newcommand{\Mbf}{{\ensuremath{\mathbf{M}}}}

\newcommand{\Pbf}{{\ensuremath{\mathbf{P}}}}

\newcommand{\Sbf}{{\ensuremath{\mathbf{S}}}}
\newcommand{\Tbf}{{\ensuremath{\mathbf{T}}}}

\newcommand{\Phbf}{{\ensuremath{\mathbf{\Phi}}}}
\newcommand{\Psbf}{{\ensuremath{\mathbf{\Psi}}}}

\newcommand{\defing}[1]{\textbf{\emph{\mathversion{bold}#1}}}

\newcommand{\R}{\ensuremath{\mathbb{R}}}

\newcommand{\Cx}{\ensuremath{\mathbb{C}}}

\renewcommand{\geq}{\geqslant}

\renewcommand{\leq}{\leqslant}

\newcounter{mcnt}

\newcounter{wordcnt}

\begin{document}

\title[Singularising quantum channels preserving kappa-additivity]{Singularising quantum channels \\ preserving kappa-additivity}

\author{Sviatoslav V. Dzhenzher}

\begin{abstract}
    We construct quantum channels by averaging unitary shift conjugations with respect to $\kappa$-additive probability measures on measurable abelian groups.
    The channels act on the duals of operator algebras generated by shifts and bounded measurable multiplication operators.
    They preserve $\kappa$-additivity of quantum states and admit a weak-$*$ integral representation on this class.
    If the averaging measure vanishes on singletons, every output state is singular.
    In the appendix, we also establish the corresponding Fubini's theorem for $\kappa$-additive measures.
\end{abstract}

\thanks{\hspace{-5mm}
S.\,V. Dzhenzher: sdjenjer@yandex.ru. orcid: 0009-0008-3513-4312
\\
Moscow Institute of Physics and Technology 141701, Institutskii lane, 9, Dolgoprudny, Russia}

\maketitle
\thispagestyle{empty}

\noindent \emph{Keywords:} Quantum channels; singular states; $\kappa$-additive measures; $\kappa$-additive states; averaging of shifts; Fubini's theorem.

\vspace{3mm}
\noindent \emph{MSC 2020:}
46L30, 
81P47, 
28A35. 

\section{Introduction}

Quantum channels describe transformations of quantum states compatible with positivity, normalisation, and the inclusion of auxiliary systems.
In the usual infinite-dimensional formulation, a channel acts on trace-class operators, while its adjoint is a normal unital completely positive map on the algebra of bounded operators. The theory of completely positive maps and its application to quantum information provide the standard framework
for this description \cite{Stinespring-1955, Kholevo-qsys-chan-inf}.
For a concrete operator algebra, however, positive normalised functionals need not be represented by density operators. Dynamics involving singular states therefore requires a formulation on the full state space of the chosen algebra.

Integration and averaging enter quantum channel theory in several ways.
Continuous ensembles are described by probability measures on spaces of density operators \cite{Kholevo-Shirokov-06}, and translation covariance
plays a central role in the structure of quantum Markov equations \cite{Kholevo1995}.
Extensions of channels to non-normal input states have also been studied; in particular, Shirokov obtained a criterion for an extension that is continuous from the weak-$*$ topology to the trace norm \cite{Shirokov-2009}. Such an extension still has normal output states.
The problem considered here instead concerns channels whose outputs annihilate the compact operators.

Integral representations of general states using finitely additive measures were developed by Amosov and Sakbaev \cite{Amosov-Sakbaev-2013,Amosov-Sakbaev-2016}. The role of the observable algebra in dynamics extending beyond vector states was investigated by Volovich and Sakbaev \cite{Volovich-Sakbaev-2018}. A related line of work constructs quantum evolutions by averaging random shifts.
Sakbaev studied shift-invariant measures and averaged random walks in a Hilbert space \cite{Sakbaev2017}; Busovikov, Pechen, and Sakbaev investigated Weyl representations, Gaussian averages of coordinate and momentum shifts, and limits leading to an infinite-dimensional quantum oscillator \cite{Busovikov-Pechen-Sakbaev-2024}. Measure-based constructions also arise in perturbations of quantum dynamical semigroups by covariant
operator-valued measures \cite{Amosov-Baitenov-Pechen-2023}.

Averaging can substantially change the properties of a quantum evolution.
For example, mixtures of dynamical maps can change their divisibility properties \cite{Filippov-Piilo-Maniscalco-Ziman-2017}.
In infinite-dimensional settings, another possibility is the appearance of singular states.
Amosov and Sakbaev studied the passage from vector states to general states under averages of random shifts \cite{Amosov-Sakbaev-2025}.
Busovikov and Sakbaev established a dichotomy for Gaussian shift averages between strongly continuous quantum dynamics and the destruction of normality, depending on the covariance operator \cite{Busovikov-Sakbaev-26}.
Singularisation under averages of operator random walks was also studied in \cite{DzhenzherDzhenzherSakbaev26}. These results motivate the question
of which additivity properties of states can survive singularisation.

The connection between additivity and singular states has a set-theoretic aspect. Blecher and Weaver proved that the existence of singular countably additive states on $\Bcal(\ell^2(\kappa))$ is equivalent to the existence of suitable real-valued measures on the full power set of a cardinal $\kappa$, and obtained corresponding characterisations involving pure states and measurable cardinals \cite{Blecher-Weaver}.
Their results also include an integration lemma for uncountable families of non-negative functions. More recently, Blecher and Labuschagne studied countably additive states and sequential normality on von Neumann algebras \cite{Blecher-Labuschagne}.
The distinction between the full operator algebra and a specified algebra of observables is essential when relating these results to the present construction.

In \cite{Dzhenzher26-qsing}, singularising channels preserving countable additivity were considered in the setting of quantum measurable cardinals. The full-dual formulation of quantum channels and measurability questions for averages of unitary representations were studied in \cite{Dzhenzher26-qchan-dual}.
The extension theorem in \cite{Dzhenzher26-kappa-add} provides a separate measure-theoretic ingredient: $\kappa$-additive measures can be considered on appropriate $\kappa$-algebras without assuming that their domain is the whole power set.
This raises a further question. Can an averaging construction be carried out on such a measurable domain while preserving the corresponding additivity of quantum states?
We address this question for shifts on an abelian group.

The structure of the article is as follows.
Section~\ref{s:states} treats normal and singular functionals on concrete $C^*$-algebras containing the compact operators.
Section~\ref{s:qch} introduces the measurable averaging construction, with Theorem~\ref{t:ch} being the main result.
Section~\ref{s:proofs} proves the main theorem.
Section~\ref{s:concl} discusses further questions.
Section~\ref{s:appx} (Appendix) develops the product measure and Fubini theorem for $\kappa$-additive measures; these results supply the section-measurability and interchange arguments used in the operator construction.

\section{Quantum states on $C^*$-algebras}\label{s:states}


Let $X$ be an infinite set.
Let \(\Hcal := \ell_2(X)\) be a Hilbert space, and \(\{e_x\}\) be an orthonormal basis.
Let us denote by \(\Bcal(\Hcal)\) the von Neumann algebra of linear bounded operators \(\Hcal\to\Hcal\), and by \(\Kcal(\Hcal)\subset\Bcal(\Hcal)\) the ideal of compact operators.
Let \(\Ccal\) be a unital $C^*$-algebra such that
\[
    \Kcal(\Hcal) \subset \Ccal \subset \Bcal(\Hcal).
\]
By unitality we mean that the \emph{identity operator} $\Ibf$ lies in \(\Ccal\).

Recall that for a family \(\{c_i\}_{i\in I}\) of non-negative real numbers,
\[
    \sum_{i\in I} c_i = \sup\left\{\sum_{i \in F} c_i\;:\; F\subset I,\,\abs{F} < \infty\right\}.
\]
The sum is finite only if the collection of strictly positive reals is at most countable.
If \(\{c_i\}\) is a family of complex numbers, then
\[
    \sum_{i\in I}c_i = c
\]
means that for any \(\varepsilon>0\) there exists such finite \(F_0\subset I\) that for any finite \(F \supset F_0\),
\[
    \abs{c - \sum_{i\in F}c_i} < \varepsilon.
\]

For \(\Abf\in\Bcal(\Hcal)\), its \emph{trace} is defined by
\[
    \Tr \Abf := \sum_{x \in X} (e_x, \Abf e_x).
\]
We say that \(\Tbf\in\Bcal(\Hcal)\) is a \emph{trace class operator} if \(\Tr\abs{\Tbf} < +\infty.\)
We denote this class by \(\Tcal_1(\Hcal)\).

\begin{itemize}
\item 
For a functional \(\rho\in\Ccal^*\), we denote \(\braket{\rho,\Abf} := \rho(\Abf)\).

\item 
A functional \(\rho\in\Ccal^*\) is called a \defing{quantum state} if it is positive and \(\braket{\rho,\Ibf} = 1\).

\item 
A functional \(\rho\in\Ccal^*\) is said to be \defing{normal} if it can be identified with \(\rho\in\Tcal_1(\Hcal)\) by
\[
    \braket{\rho, \Abf} = \Tr (\rho\Abf).
\]
Note that a positive \(\rho\) is normal if it is \emph{completely additive}, which means the additivity on arbitrary families of pairwise orthogonal projectors.

\item
A functional \(\rho\in\Ccal^*\) is said to be \defing{singular} if it vanishes on all compact operators.
For a positive \(\rho\), it is equivalent to the fact that if \(\rho \geq \omega\) for some positive normal \(\omega\), then \(\omega=0\).

\item A functional \(\rho\in\Ccal^*\) is said to be \defing{$\kappa$-additive} if
\[
    \Braket{\rho, \sum_{i\in I}\Pbf_i} = \sum_{i\in I}\Braket{\rho, \Pbf_i}
\]
for any family of pairwise orthogonal projectors \(\Pbf_i\) with \(\abs{I}\leq \kappa\).
Here we follow the notation from \cite{Dzhenzher26-kappa-add} and partially from \cite{Blecher-Weaver}, but not from \cite{Jech-p1-ch10}.
\end{itemize}

The following theorem is well known \cite{Takesaki1979, Farah-Wofsey} for some particular cases, for example, when \(\Ccal\) is a von Neumann algebra. It was implicitly used in \cite{DzhenzherDzhenzherSakbaev26}.

\begin{theorem}
    For any \(\rho\in\Ccal^*\), there exists a unique decomposition 
    \[
        \rho = \rho_n + \rho_s,
    \]
    where \(\rho_n\) and \(\rho_s\) are the normal and singular parts.
    Moreover, if \(\rho\) is a quantum state, then there exists a unique convex decomposition
    \[
        \rho = \lambda\rho_n + (1-\lambda)\rho_s,
    \]
    where \(\lambda\in[\,0,1\,]\), and both \(\rho_n,\rho_s\) are quantum states.
\end{theorem}

\begin{proof}
    The uniqueness follows since \(\rho_n=\rho_s\) implies \(\rho_n=\rho_s=0\).
    
    Since \(\Kcal(\Hcal)^* = \Tcal_1(\Hcal)\), there exists \(\Tbf\in\Tcal_1(\Hcal)\) such that
    \[
        \braket{\rho,\Kbf} = \Tr(\Tbf\Kbf)
        \quad\text{for all}\quad \Kbf\in\Kcal(\Hcal).
    \]
    So, define
    \[
        \braket{\rho_n, \Abf} := \Tr (\Tbf\Abf)
        \quad\text{for}\quad \Abf\in\Ccal,
        \qquad \rho_s := \rho - \rho_n.
    \]
    It is clear from the definition that \(\rho_n\) is normal.
    And \(\rho_s\) is singular since
    \[
        \braket{\rho,\Kbf} - \braket{\rho_n,\Kbf} = \Tr(\Tbf\Kbf) - \Tr(\Tbf\Kbf) = 0.
    \]

    It remains to show that if \(\rho\) is positive, then so are \(\rho_n\) and \(\rho_s\). After this, \(\lambda\) can be chosen by rescaling, since
    \[
        1 = \braket{\rho,\Ibf} = \braket{\rho_n, \Ibf} + \braket{\rho_s, \Ibf}.
    \]
    It is clear that \(\Tbf\) should be positive if \(\rho\) is such.
    Hence, \(\rho_n\) is positive.
    Let us show that \(\rho \geq \rho_n\), which will imply \(\rho_s\geq 0\).
    So, let \(\Abf \in\Ccal\) be positive.
    For any finite-dimensional projector \(\Pbf\), we have
    \[
        0 \leq \sqrt{\Abf}\Pbf\sqrt{\Abf} \leq \Abf,
    \]
    and \(\sqrt{\Abf}\Pbf\sqrt{\Abf}\) is a compact operator.
    Hence,
    \[
        \braket{\rho, \Abf} \geq \Braket{\rho, \sqrt{\Abf}\Pbf\sqrt{\Abf}} = \Tr \left(\Tbf\sqrt{\Abf}\Pbf\sqrt{\Abf}\right).
    \]
    With \(\Pbf\) converging to \(\Ibf\), we obtain
    \[
        \braket{\rho, \Abf} \geq \Tr (\Tbf\Abf)  = \braket{\rho_n,\Abf}. \qedhere
    \]
\end{proof}

\section{Quantum channels and measurability}\label{s:qch}

Now, let us additionally assume that \(X\) is an abelian group \((X,+)\).

Let $\kappa$ be an infinite cardinal.
Let \(\Acal\) be a $\kappa$-algebra of subsets of \(X \in \Acal\) that is invariant under shifts and contains all singletons.
Let \(\mu\colon\Acal\to[\,0,1\,]\) be a probability $\kappa$-additive measure (not necessarily invariant under shifts).
Recall that $\kappa$-additivity means that
\[
    \mu\left(\bigsqcup_{i\in I} A_i\right) = \sum_{i\in I}\mu(A_i)
\]
for \(A_i\in \Acal\) and \(\abs{I}\leq\kappa\).

For any \(x \in X\), define the unitary shift operator \(\Sbf_x\colon\Hcal\to\Hcal\) by
\[
    \Sbf_xu(y) := u(x+y).
\]

For a bounded \(\Acal\)-measurable function \(f\colon X\to\Cx\), define the bounded multiplication operator \(\Mbf_f\colon\Hcal\to\Hcal\) by
\[
    \Mbf_f u(x) := f(x)u(x).
\]

Following \cite{DzhenzherDzhenzherSakbaev26}, we present the unital $C^*$-algebra \(\Ccal \subset \Bcal(\Hcal)\) generated by all shift operators \(\Sbf_x\) and multiplication operators \(\Mbf_g\).
It is clear that \(\Ccal\supset\Kcal(\Hcal)\), since a projector on \(x\in X\) is a multiplication by the indicator function of \(\{x\}\).
Note that by the analogue of CCR
\[
    \Mbf_f\Sbf_x u(y) = f(y)u(x+y) = \Sbf_x\Mbf_{\Sbf_x^* f} u(y),
\]
we obtain that this $C^*$-algebra is the span (in the norm topology) of linear combinations of the form \(\Sbf_x\Mbf_f\).

Our main goal is to define the quantum channel \(\Phbf_\mu\colon\Ccal^*\to\Ccal^*\) by the weak-$*$ integral:
\[
    \Phbf_\mu\rho = \int \Sbf_x\rho\Sbf_x^*\,d\mu(x),
\]
meaning
\[
    \Braket{\Phbf_\mu\rho, \Abf} =
    \int \Braket{\Sbf_x\rho\Sbf_x^*, \Abf}\,d\mu(x) =
    \int \Braket{\rho, \Sbf_x^*\Abf\Sbf_x}\,d\mu(x).
\]
If we prove that the integrand is a measurable function, then it will be automatically integrable, since
\[
    \abs{\Braket{\rho, \Sbf_x^*\Abf\Sbf_x}} \leq \norm{\rho}\norm{\Abf},
\]
and the measure is finite.
If \(\mu\) were defined on all subsets of $X$, as in \cite{Dzhenzher26-qsing}, then no problems with measurability would arise.
See, for example, in \cite{Dzhenzher26-qchan-dual}, how the integrand can be non-measurable for a Borel measure.
There, the measurability was cured with the continuity of a unitary representation in the norm topology.
Here we suggest a measure-theoretical approach.

Denote by \(\Acal \otimes_\kappa \Acal\) the $\kappa$-algebra generated by the measurable rectangles \(A\times B\) for \(A,B\in\Acal\).
Suppose additionally that the group operation \(+\colon X\times X\to X\) is \((\Acal\otimes_\kappa\Acal) |\Acal\)-measurable, meaning
\[
    +^{-1}(A) \in \Acal\otimes_\kappa\Acal
    \quad\text{for}\quad A\in\Acal.
\]
Define the \defing{quantum channel in the Heisenberg picture} \(\Psbf_\mu\colon\Ccal\to\Ccal\) by the weak integral
\[
    \Psbf_\mu(\Abf) := \int \Sbf_x^*\Abf\Sbf_x\,d\mu(x),
\]
meaning
\[
    (u, \Psbf_\mu(\Abf)v) = \int (\Sbf_x u, \Abf\Sbf_x v)\,d\mu(x).
\]
After that, define \defing{quantum channel in the Schr\"odinger picture} \(\Phbf_\mu\colon\Ccal^*\to\Ccal^*\) by
\[
    \Phbf_\mu := \Psbf_\mu^*,
\]
which means
\[
    \Braket{\Phbf_\mu(\rho), \Abf} := \Braket{\rho, \Psbf_\mu(\Abf)}.
\]
We prove the correctness of both definitions and the following theorem in the next section.

\begin{theorem}\label{t:ch}
    Let \(\Acal\) be a $\kappa$-algebra of subsets of \(X \in \Acal\) that is invariant under shifts and contains all singletons.
    Let \(+\) be \((\Acal\otimes_\kappa\Acal) |\Acal\)-measurable.
    Then the channel \(\Psbf_\mu\) is well defined, and the channel \(\Phbf_\mu\) preserves $\kappa$-additivity of quantum states.
    Moreover, the following integral representation holds for a $\kappa$-additive quantum state $\rho$:
    \[
        \Phbf_\mu\rho = \int \Sbf_x\rho\Sbf_x^*\,d\mu(x).
    \]
    If additionally $\mu$ vanishes on singletons (and hence on all finite subsets), then \(\Phbf_\mu\) is singularising.
\end{theorem}

Here, the \defing{preservation of $\kappa$-additivity} means that \(\Phbf_\mu\rho\) is $\kappa$-additive if $\rho$ is such.
The \defing{singularisation} means that the output functionals are always singular.
See examples of appropriate measures, for example, in \cite{Dzhenzher26-kappa-add}.
Another simple example is \(\kappa=\aleph_0\) and \(\Acal\) being the Borel $\sigma$-algebra on $\R$.
Thus, Theorem~\ref{t:ch} resolves problems with measurability that arose in \cite{DzhenzherDzhenzherSakbaev26} in another way: instead of constructing the integral on pure states and then extending it to the entire state space, the theorem provides a more measure-theoretical approach.

Note that if \(\abs{X} \leq \kappa\), then $\kappa$-additivity is equivalent to complete additivity, and thus the channel $\Phbf_\mu$ preserves the normality of quantum states.
In this case, $\mu$ cannot vanish on singletons, and the channel cannot be singularising.

\section{Proofs}\label{s:proofs}

Recall that in this entire section,

\begin{itemize}
    \item \((X,+)\) is an abelian group,
    \item $\Acal$ is a $\kappa$-algebra of subsets of \(X \in \Acal\) that is invariant under shifts and contains all singletons,
    \item $\mu$ is a $\kappa$-additive probability measure on $\Acal$,
    \item \(+\) is \((\Acal\otimes_\kappa\Acal) |\Acal\)-measurable,
    \item $\Hcal = \ell_2(X)$,
    \item $\Ccal$ is generated by shifts and multiplications.
\end{itemize}

\begin{proof}[Proof of correctness of the definition of $\Psbf$]
    It is sufficient to show that for fixed \(u,v\in\Hcal\) and \(\Abf\in\Ccal\), the integrand \((\Sbf_x u, \Abf\Sbf_x v)\) is integrable.
    Since \(\norm{(\Sbf_x u, \Abf\Sbf_x v)} \leq \norm{u}\norm{v}\norm{\Abf}\) is bounded, it is sufficient to prove measurability.
    Since \(\Ccal\) is generated by \(\Sbf_y\) and \(\Mbf_f\), it is sufficient to prove measurability for \(\Abf\) being either a shift or a multiplication.

    Let \(\Abf = \Sbf_y\) for some \(y\in X\).
    Then
    \[
        (\Sbf_x u, \Abf\Sbf_x v) = (\Sbf_x u, \Sbf_{x+y} v) = (u,\Sbf_y v)
    \]
    is constant, and hence integrable.
    Strictly speaking, this case was clear without any $u,v$ since
    \[
        \Psbf_\mu(\Sbf_y) = \int \Sbf_{-x}\Sbf_y\Sbf_x\,d\mu(x) = \Sbf_y\int \,d\mu = \Sbf_y.
    \]

    Let \(\Abf = \Mbf_f\) for some bounded \(\Acal\)-measurable \(f\colon X\to\Cx\).
    Then
    \[
        (\Sbf_x u, \Abf\Sbf_x v) = \sum_{y\in X} \overline{u(x+y)}f(y)v(x+y),
    \]
    which is measurable since the addition is measurable, and since measurability is preserved under countable summation.
    Thus, analogously to \cite{DzhenzherDzhenzherSakbaev26}, we obtain
    \[
        \Psbf_\mu(\Mbf_f) = \Mbf_g
        \quad\text{for}\quad
        g = \int \Sbf_x^* f\,d\mu(x). \qedhere
    \]
\end{proof}

\begin{proof}[Proof of the integral representation in Theorem~\ref{t:ch}]
    We need to show that
    \[
        \Braket{\rho, \int \Sbf_x^*\Abf\Sbf_x\,d\mu(x)} = \int\Braket{\rho, \Sbf_x^*\Abf\Sbf_x}\,d\mu(x).
    \]
    For \(\Abf = \Sbf_y\) it is clear, so we need to check only \(\Abf = \Mbf_f\) for some bounded measurable $f$.
    
    Define the $\kappa$-additive probability measure \(\nu\colon\Acal\to[\,0,1\,]\) by
    \[
        \nu(E) := \Braket{\rho, \Pbf_E},
    \]
    where \(\Pbf_E\) is the projector onto \(\ell_2(E)\).
    Then
    \[
        \Braket{\rho, \Mbf_f} = \int f(y)\,d\nu(y),
    \]
    and, as it was shown,
    \[
        \Psi_\mu(\Mbf_f) = \int \Sbf_x^*\Mbf_f\Sbf_x\,d\mu(x) = \Mbf_g
        \quad\text{for}\quad
        g = \int \Sbf_x^* f\,d\mu(x).
    \]
    Hence
    \[
        \Braket{\rho, \int \Sbf_x^*\Mbf_f\Sbf_x\,d\mu(x)} =
        \int g(y)\,d\nu(y) =
        \int f(y-x)\,d\mu(x)\,d\nu(y) =
        \int\Braket{\rho, \Sbf_x^*\Abf\Sbf_x}\,d\mu(x),
    \]
    where the last equality is by Fubini's theorem for $\kappa$-additive measures; see Theorem~\ref{t:fubini} and Section~\ref{s:appx} (Appendix) for details.
\end{proof}

The following lemma, which is the $\kappa$-additive generalisation of the Beppo Levi monotone convergence theorem, is known \cite[Lemma~3.2]{Blecher-Weaver}.
It was proved there by transfinite induction. Here we provide another proof.

\begin{lemma}[Levi, Blecher, Weaver]\label{l:levi-series}
    Let \(\{f_i\}_{i\in I}\) be a family of non-negative bounded measurable functions \(X\to\Cx\), and \(\abs{I}\leq\kappa\).
    Then
    \[
        \sum_{i\in I}\int f_i\,d\mu = \int \sum_{i\in I}f_i\,d\mu.
    \]
\end{lemma}

\begin{proof}
    It is clear that
    \[
        \sum_{i\in I}\int f_i\,d\mu \leq \int \sum_{i\in I}f_i\,d\mu.
    \]
    So, if
    \[
        \sum_{i\in I}\int f_i\,d\mu = +\infty,
    \]
    then the lemma is proved.
    Suppose that
    \[
        \sum_{i\in I}\int f_i\,d\mu < +\infty.
    \]
    Then the family \(I_+\) of \(i\in I\) such that \(\int f_i\,d\mu > 0\) is at most countable.
    For \(i\notin I_+\) we have \(\int f_i=0\), which with the non-negativity of $f_i$ implies \(f_i = 0\) almost everywhere.
    Since \(\mu\) is $\kappa$-additive and \(\abs{I}\leq\kappa\), we obtain that simultaneously all \(f_i=0\) for \(i\notin I_+\) almost everywhere.
    Then, ignoring \(f_i\) not from \(I_+\), we obtain that
    \[
        \sum_{i\in I}\int f_i\,d\mu = \sum_{i\in I_+}\int f_i\,d\mu = \int\sum_{i\in I_+} f_i\,d\mu = \int\sum_{i\in I} f_i\,d\mu,
    \]
    where the middle equality is by the classical Levi monotone convergence theorem for series.
\end{proof}

\begin{proof}[Proof of $\kappa$-additivity preservation in Theorem~\ref{t:ch}]
    Let \(\{\Pbf_i\}_{i\in I}\) be pairwise orthogonal and \(\abs{I}\leq\kappa\).
    Then
    \begin{multline*}
        \Braket{\Phbf_\mu\rho, \sum_{i\in I}\Pbf_i} =
        \int\Braket{\rho, \sum_{i\in I}\Sbf_x^*\Pbf_i\Sbf_x}\,d\mu(x) =
        \int\sum_{i\in I}\Braket{\rho, \Sbf_x^*\Pbf_i\Sbf_x}\,d\mu(x) = \\ =
        \sum_{i\in I}\int\Braket{\rho, \Sbf_x^*\Pbf_i\Sbf_x}\,d\mu(x) =
        \sum_{i\in I}\Braket{\Phbf_\mu\rho, \Pbf_i},
    \end{multline*}
    where
    \begin{itemize}
        \item the first and the last equalities are by the already proven integral representation for \(\Phbf_\mu\),
        \item the second equality is by $\kappa$-additivity of $\rho$, and
        \item the third equality is by Lemma~\ref{l:levi-series}. \qedhere
    \end{itemize}
\end{proof}

\begin{proof}[Proof of singularisation in Theorem~\ref{t:ch}]
    The singularisation means that for any \(\rho\in\Ccal^*\) and any \(\Kbf\in\Kcal(\Hcal)\),
    \[
        \Braket{\rho, \Psbf_\mu\Kbf} = \Braket{\Phbf_\mu\rho, \Kbf} = 0.
    \]
    For a fixed $\Kbf$, this means that \(\Psbf_\mu\Kbf = 0\).
    It is sufficient to check this for finite-dimensional projectors.
    So, let \(F\subset X\) be finite.
    Then the set of $x$ such that \((\Sbf_x u, \Pbf_F\Sbf_x v) \neq 0\) is at most countable.
    Hence
    \[
        (u,\Psbf_\mu\Pbf_F v) = \int (\Sbf_x u, \Pbf_F\Sbf_x v)\,d\mu(x) = 0. \qedhere
    \]
\end{proof}

\section{Conclusion and further research}\label{s:concl}

We have constructed singularising quantum channels from $\kappa$-additive probabilities on measurable abelian groups.
The construction uses the algebra generated by shifts and measurable multiplication operators, together with the joint measurability of addition.
On this algebra, weak operator averaging produces a unital completely positive map whose adjoint acts on the full dual. Its restriction to $\kappa$-additive quantum states is represented by weak-$*$ integration and preserves $\kappa$-additivity.
When the averaging measure vanishes on singletons, the averaged map annihilates every compact operator, so all output states are singular.

The measure-theoretic and operator-algebraic parts of the construction are linked by the choice of measurable observables.
The $\kappa$-monotone class theorem provides measurability of section measures and section integrals, and hence Fubini's theorem on the generated $\kappa$-product algebra. These results make it possible to work with measures on specified $\kappa$-algebras while controlling the matrix coefficients and scalar functionals needed for quantum averaging.

An important next step is to extend the integral representation of Theorem~\ref{t:ch} to the entire space of quantum channels.
A second direction is to study convolution families of such measures and the resulting channel semigroups, including their continuity and long-time behaviour. For example, it may be interesting to focus on the criteria when the channels become idempotent depending on $\mu$.
Finally, one may ask when a channel or an output state admits an extension from $\Ccal$ to $\Bcal(\ell^2(X))$ that preserves the same
additivity.
This extension problem connects the present measurable construction with the constraints identified by the theory of quantum measurable cardinals.

\section{Appendix: $\kappa$-additive version of Fubini's theorem}\label{s:appx}

Here we construct the product measure and prove Fubini's theorem for $\kappa$-additive measures.
The construction is analogous to the classical one; see, for example, \cite{Halmos}.

We say that a class \(\Mcal\) of sets is \emph{$\kappa$-monotone} if it is closed under monotone $\kappa$-intersections and $\kappa$-unions.
Strictly speaking, let \(\gamma\) be an ordinal such that \(\abs{\gamma} \leq \kappa\).
Let \(\{E_\alpha\}_{\alpha < \gamma}\) be a family of sets from \(\Mcal\) such that
\[
    \alpha < \beta \implies E_\alpha \subset E_\beta.
\]
Then \(\bigcup_{\alpha < \gamma} E_\alpha \in \Mcal\).
The closedness under $\kappa$ monotone intersections is defined analogously.

\begin{lemma}[Monotone class]\label{l:monotone}
    Let \(\Acal\) be an algebra of subsets of $X$.
    Then \(\kappa(\Acal)\) coincides with the minimal $\kappa$-monotone class over $\Acal$.
\end{lemma}

\begin{proof}
    Let \(\Mcal\) be the monotone class generated by $\Acal$.
    It is clear that \(\kappa(\Acal) \supset \Mcal\), so we need to prove that \(\kappa(\Acal) \subset \Mcal\).
    In other words, that \(\Mcal\) is a $\kappa$-algebra.

    Denote
    \[
        \Mcal_0 := \{A \in \Mcal : X\setminus A \in \Mcal\}.
    \]
    By De Morgan's laws, $\Mcal_0$ is a $\kappa$-monotone class over $\Acal$. Hence, \(\Mcal_0 = \Mcal\), which means that $\Mcal$ is closed under complement.

    For any \(A\in \Mcal\), denote by
    \[
        \Mcal_A := \{B \in \Mcal: A\cap B \in \Mcal\}
    \]
    the class of sets that are `well intersected with $A$'.
    It is clear that for any \(A \in \Acal\), the class \(\Mcal_A\) is a $\kappa$-monotone class over $\Acal$, and thus \(\Mcal = \Mcal_A\).
    This means that all elements from $\Mcal$ intersect well with elements of $\Acal$.
    Thus, analogously, for any \(A' \in \Mcal\) we have \(\Mcal_{A'} = \Mcal\), which means that $\Mcal$ is closed under finite intersections.

    So, we have just proved that $\Mcal$ is an algebra.
    Now, let \(\{E_i\}_{i \in I}\) be from \(\Mcal\), and \(\abs{I}\leq \kappa\).
    Well order $I$ so that it is the least ordinal of cardinality $\kappa$.
    Taking
    \[
        U_0 = \varnothing,\quad
        U_{\alpha+1} = U_\alpha \cup E_\alpha,\quad
        U_\beta = \bigcup_{\alpha < \beta} U_\alpha,
    \]
    by transfinite induction, it is easy to show that \(U_\alpha \in \Mcal\) for any \(\alpha \leq \gamma\).
    In particular,
    \[
        U_\gamma = \bigcup_{\alpha < \gamma} E_\alpha \in \Mcal.\qedhere
    \]
\end{proof}

Analogously, we say that a measure \(\mu\) is \emph{$\kappa$-continuous from above} if
\[
    \mu\left(\bigcap_{\alpha < \gamma} E_\alpha\right) = \inf_{\alpha < \gamma} \mu(E_\alpha),
\]
and the definition of \emph{$\kappa$-continuity from below} is analogous.

\begin{proposition}
    Any $\kappa$-additive finite measure of a $\kappa$-algebra is $\kappa$-continuous.
\end{proposition}

\begin{proof}
    The proof is analogous to the classical proof and the last step of the proof of Lemma~\ref{l:monotone}.
\end{proof}

Below, we will need the following analogue of the Levi monotone convergence theorem, but for functions instead of series.

\begin{lemma}\label{l:levi}
    Let $\mu$ be a finite $\kappa$-additive measure on a $\kappa$-algebra.
    Let \(f_\alpha\geq 0\) be measurable and monotone,
    \[
        f := \sup_{\alpha < \gamma} f_\alpha,
        \quad \abs{\gamma} \leq \kappa.
    \]
    Then $f$ is measurable and
    \[
        \int f\,d\mu = \sup_{\alpha < \gamma} \int f_\alpha\,d\mu.
    \]
\end{lemma}

\begin{proof}
    First, note that $f$ is measurable since
    \[
        \{f > t\} = \bigcup_{\alpha < \gamma} \{f_\alpha > t\}
    \]
    lies in the $\kappa$-algebra.

    It is clear that
    \[
        \int f\,d\mu \geq \sup_{\alpha < \gamma} \int f_\alpha\,d\mu,
    \]
    so we need to show the other inequality.
    Let \(\phi \leq f\) be a non-negative measurable simple function, taking values \(0,\phi_1,\ldots,\phi_n\).
    Fix \(0<c<1\) and consider the sets
    \[
        B_{j,\alpha} := \{\phi=\phi_j\}\cap \{f_\alpha > c\phi_j\},
        \quad j=1,\ldots,n, \quad \alpha < \gamma
    \]
    that are monotone in $\alpha$.
    Since $c\phi_j$ is strictly less than $f$ on \(\{\phi=\phi_j\}\),
    \[
        \bigcup_{\alpha<\gamma} B_{j,\alpha} = \{\phi=\phi_j\}.
    \]
    Hence
    \[
        \sup_{\alpha<\gamma}\mu(B_{j,\alpha}) = \mu\{\phi=\phi_j\},
    \]
    which together with
    \[
        \int f_\alpha\,d\mu \geq c \sum_{j=1}^n\phi_j \mu(B_{j,\alpha})
    \]
    implies
    \[
        \sup_{\alpha < \gamma} \int f_\alpha\,d\mu \geq c \sum_{j=1}^n\phi_j \mu\{\phi=\phi_j\} = c\int\phi\,d\mu.
    \]
    Since $c$ is arbitrary, we have
    \[
        \sup_{\alpha < \gamma} \int f_\alpha\,d\mu \geq \int\phi\,d\mu.
    \]
    It remains to use that \(\int f\,d\mu = \sup_{\phi \leq f}\int \phi\,d\mu\).
\end{proof}

\begin{lemma}\label{l:extend}
    Let \(\mu\colon\Scal\to[\,0,1\,]\) be a $\kappa$-additive measure on a semi-algebra $\Scal$.
    Then it can be extended to the generated algebra $\Acal$ in a unique way, and the $\kappa$-additivity is preserved.
\end{lemma}

\begin{proof}
    It is well known that the generated algebra consists of subsets of the form
    \[
        \bigsqcup_{k=1}^n A_k,
    \]
    where \(A_k\in\Scal\).
    It is clear that
    \[
        \mu\left(\bigsqcup_{k=1}^n A_k\right) := \sum_{k=1}^n \mu(A_k)
    \]
    is well defined and finitely additive.
    So, suppose that \(B = \bigsqcup_{i \in I}B_i\) for \(\abs{I}\leq\kappa\) and \(B_i\) are from generated algebra.
    Take decompositions
    \[
        B = \bigsqcup_{k=1}^n A_k,
        \quad
        B_i = \bigsqcup_{m=1}^{n_i} A_{i,m}.
    \]
    Then
    \[
        \mu(B) = \sum_{k=1}^n \mu(A_k) =
        \sum_{k=1}^n\sum_{i\in I}\sum_{m=1}^{n_i} \mu(A_k \cap A_{i,m}) = \sum_{i\in I}\sum_{m=1}^{n_i}\mu(A_{i,m}) = \sum_{i\in I}\mu(B_i),
    \]
    where in the third equality we used the fact that \(\mu(B) < +\infty\), hence in the sum there is only at most countable family of non-zero summands, and so we can change the order of summation by the absolute convergence of series.
\end{proof}

Now, we turn to the construction of the product-measure.
Let \((X,\Acal_X,\mu)\) and \((Y,\Acal_Y,\nu)\) be two measurable spaces with probability $\kappa$-additive measures on $\kappa$-algebras.
It is clear that \(\Acal_X \times \Acal_Y\) is a semi-algebra of subsets of \(X\times Y\).
Define the pre-measure on it by
\[
    \mu \times \nu(A\times B) := \mu(A)\nu(B).
\]

\begin{figure}
    \centering
    \begin{tikzpicture}
    \def \ux {5};
    \def \uy {5};
    
    \draw[black, very thick] (0,0) rectangle (\ux,\ux);
    \draw[black, thick] (0,0) rectangle (2,1);
    \draw[black, thick] (1,1) rectangle (3,3);
    \draw[black, thick] (0,1) rectangle (1,4);
    \draw[black, thick] (2,0) rectangle (4,1);
    \draw[black, thick] (4,0) rectangle (5,4);
    \draw[black, thick] (0,4) rectangle (3,5);
    \draw[black, thick] (3,1) rectangle (4,5);

    \draw[<->, red, thick] (0,1.5) -- (\ux,1.5);
    \draw[<->, red, thick] (0,1.7) -- (1,1.7);
    \draw[<->, red, thick] (1,1.7) -- (3,1.7);
    \draw[<->, red, thick] (3,1.7) -- (4,1.7);
    \draw[<->, red, thick] (4,1.7) -- (5,1.7);
    
    \node (S) at (-0.3,\uy/2) {$A$};
    \node (T) at (\ux/2,\uy+0.3) {$B$};
    \end{tikzpicture}
    \caption{Check of the $\kappa$-additivity on a semi-algebra of sets}
    \label{fig:prod-m-levi}
\end{figure}
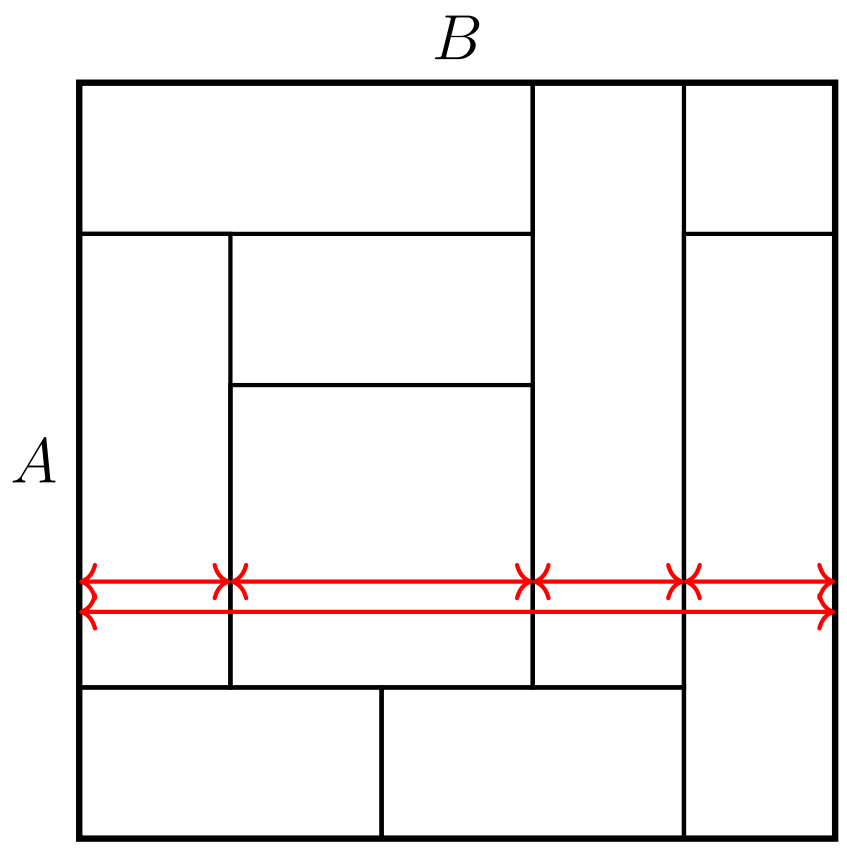

\begin{lemma}
    The pre-measure \(\mu \times \nu(A\times B) := \mu(A)\nu(B)\) is $\kappa$-additive.
\end{lemma}

\begin{proof}
    Suppose \(A\times B = \bigsqcup_{i\in I} A_i\times B_i\) for \(\abs{I}\leq\kappa\) and \(A_i\in \Acal_X\), \(B_i\in\Acal_Y\).
    Consider non-negative measurable functions
    \[
        f_i := \Ical(A_i)\nu(B_i)\colon X\to\Cx,
    \]
    where \(\Ical\) is the indicator-function.
    Then (see Fig.~\ref{fig:prod-m-levi})
    \[
        \sum_{i\in I} f_i = \Ical(A)\nu(B).
    \]
    Hence, by Lemma~\ref{l:levi-series},
    \[
        \sum_{i\in I}\mu\times \nu(A_i\times B_i) =
        \sum_{i\in I} \int f_i\,d\mu =
        \int_A \nu(B)\,d\mu = \mu\times\nu(A\times B). \qedhere
    \]
\end{proof}

For \(A\subset X\times Y\), we denote by \(A_x\) and \(A^y\) the corresponding sections.
Analogously, for \(f\colon X\times Y\to\Cx\), we denote by \(f_x\colon Y\to\Cx\) and \(f^y\colon X\to\Cx\) the corresponding sections.
We also denote \(\{f\in C\} := f^{-1}(C)\).

\begin{proposition}
    For any \(x\in X\) and any measurable set \(A\in \Acal_X\otimes_\kappa \Acal_Y\), its section is measurable: \(A_x \in \Acal_Y\).
\end{proposition}

\begin{proof}
    This follows since the set
    \[
        \{A \in \Acal_X\otimes_\kappa\Acal_Y : A_x \in\Acal_Y\}
    \]
    is a $\kappa$-algebra over \(\Acal_X\times\Acal_Y\).
\end{proof}

\begin{proposition}
    For any \(x\in X\) and any \(\Acal_X\otimes_\kappa \Acal_Y\)-measurable function, its section \(f_x\) is \(\Acal_Y\)-measurable.
\end{proposition}

\begin{proof}
    This follows from the previous proposition, since
    \[
        \{f_x \in C\} = \{f \in C\}_x. \qedhere
    \]
\end{proof}

\begin{proposition}\label{p:meas-sect}
    For any \(A \in \Acal_X\otimes_\kappa\Acal_Y\), the function \(x \mapsto \nu(A_x)\) is $\Acal_X$-measurable and $\mu$-integrable.
\end{proposition}

\begin{proof}
    Denote these functions by $f_A\colon X\to [\,0,1\,]$.
    Denote by
    \[
        \Fcal := \{A \in \Acal_X\otimes_\kappa \Acal_Y : \text{$f_A$ is measurable}\}.
    \]
    It is clear that \(\Fcal\supset \Acal_X\times\Acal_Y\).
    It is also clear that $\Fcal$  contains the algebra generated by \(\Acal_X\times\Acal_Y\), since for \(A = \bigsqcup_{n=1}^N A_n\) we have \(f_A = \sum_{n=1}^N f_{A_n}\).
    Also, by the $\kappa$-continuity of $\nu$ and by Lemma~\ref{l:levi}, the class $\Fcal$ is $\kappa$-monotone. Hence, it coincides with \(\Acal_X\otimes_\kappa \Acal_Y\).
\end{proof}

\begin{corollary}
    For any bounded \(\Acal_X\otimes_\kappa \Acal_Y\)-measurable function \(f\colon X\times Y\to\Cx\), the function
    \[
        x \mapsto \int f(x,y)\,d\nu(y)
    \]
    is $\Acal_X$-measurable and $\mu$-integrable.
\end{corollary}

\begin{proof}
    Since we can uniformly approximate $f$ by simple functions, it is sufficient to consider the case when $f$ is simple.
    For a simple $f$, this follows from the previous proposition.
\end{proof}



\begin{theorem}[Construction of the product-measure]\label{t:product-m}
    Let \(\mu\colon\Acal_X\to[\,0,1\,]\) and \(\nu\colon\Acal_Y\to[\,0,1\,]\) be $\kappa$-additive probability measures on $\kappa$-algebras $\Acal_X,\Acal_Y$.
    Define the measure \(\mu\otimes\nu\colon\Acal_X\otimes_\kappa\Acal_Y\to[\,0,1\,]\) by
    \[
        \mu\otimes\nu(A) := \int\nu(A_x)\,d\mu(x).
    \]
    Then \(\mu\otimes\nu\) is a unique $\kappa$-additive extension of $\mu\times\nu$, and
    \[
        \mu\otimes\nu(A) = \int_X \nu(A_x)\,d\mu(x) = \int_Y \mu(A^y)\,d\nu(y).
    \]
\end{theorem}

\begin{proof}
    The correctness of the definition follows by Proposition~\ref{p:meas-sect}.
    It is clear that \(\mu\otimes\nu\) extends $\mu\times\nu$.
    Its $\kappa$-additivity follows since if \(A = \bigsqcup_{i\in I}A_i\) for \(\abs{I}\leq\kappa\), then \(A_x = \bigsqcup_{i\in I} (A_i)_x\), and
    \[
        \nu(A_x) = \sum_{i\in I} \nu((A_i)_x),
    \]
    and by Lemma~\ref{l:levi-series},
    \[
        \mu\otimes\nu(A) = \sum_{i\in I} \mu\otimes\nu(A_i).
    \]

    The uniqueness of $\mu\otimes\nu$ can be proved as follows.
    Let $\eta$ be another $\kappa$-additive extension of $\mu\times\nu$.
    Then, by $\kappa$-monotonicity of all measures and by Lemma~\ref{l:extend}, the class
    \[
        \Mcal = \{A \in \Acal_X\otimes_\kappa\Acal_Y: \mu\otimes\nu(A) = \eta(A)\}
    \]
    is a monotone class over the algebra generated by $\Acal_X\times\Acal_Y$.
    Hence \(\Mcal = \Acal_X\otimes_\kappa\Acal_Y\) by Lemma~\ref{l:monotone}, and \(\mu\otimes\nu = \eta\).

    Finally, the equality
    \[
        \mu\otimes\nu(A) = \int_Y \mu(A^y)\,d\nu(y)
    \]
    holds since we can construct the measure \(\eta\) by the right integral, and by uniqueness this should coincide with $\mu\otimes\nu$.
\end{proof}

\begin{theorem}[Fubini, Tonelli]\label{t:fubini}
    Let \(\mu\colon\Acal_X\to[\,0,1\,]\) and \(\nu\colon\Acal_Y\to[\,0,1\,]\) be $\kappa$-additive probability measures on $\kappa$-algebras $\Acal_X,\Acal_Y$.
    Let \(f\colon X\times Y\to \Cx\) be a \(\mu\otimes\nu\)-integrable \(\Acal_X\otimes_\kappa\Acal_Y\)-measurable function. Then
    \[
        \int_{X\times Y} f\,d(\mu\otimes\nu) =
        \int_X d\mu(x) \int_Y f(x,y)\,d\nu(y) =
        \int_Y d\nu(y) \int_X f(x,y)\,d\mu(x).
    \]
    Moreover, if $f\geq 0$ is just measurable, then the finiteness of the second or third integral implies $\mu\otimes\nu$-integrability.
\end{theorem}

\begin{proof}
    Taking the real, imaginary, positive, and negative parts, it is sufficient to consider the case $f \geq 0$.
    By the classical argument, it is sufficient to consider bounded functions.
    Take simple measurable functions \(\phi_n\) that monotonically converge to $f$ from below.
    By Theorem~\ref{t:product-m}, the functions
    \[
        \psi_n \colon x\mapsto \int\phi_n(x,y)\,d\nu(y)
    \]
    are measurable, integrable, and
    \[
        \int \psi_n\,d\mu = \int \phi_n\,d(\mu\otimes\nu).
    \]
    By the classical Levi monotone convergence theorem \cite[Theorem~16.2]{Billingsley2012probability}, for each $x$, the values \(\psi_n(x)\) converge from below to
    \[
        \psi(x) := \int f(x,y)\,d\nu(y).
    \]
    Hence, $\psi$ is measurable and integrable.
    Thus, again by the classical Levi monotone convergence theorem \cite[Theorem~16.2]{Billingsley2012probability},
    \[
        \int\psi_n\,d\mu \xrightarrow[n\to\infty]{} \int\psi\,d\mu,
    \]
    which implies the result.
\end{proof}

\printbibliography

\end{document}